\documentclass{aa}
\usepackage{graphicx}
\usepackage{lscape}
\usepackage{epsfig}
\begin{document}

\title{Analysis of IUE spectra of helium-rich white dwarf stars
{\footnote{Partially 
based on observations at
Observat\'orio do Pico dos Dias / LNA}}
}

\author{
B. G. Castanheira\inst{1},
S. O. Kepler\inst{1},
G. Handler\inst{2}$^,$\inst{3},
D. Koester\inst{4}}

\offprints{barbara@if.ufrgs.br}
\institute{
Instituto de F\'{\i}sica, Universidade Federal do Rio Grande do Sul,
  91501-900  Porto-Alegre, RS, Brazil\\
\and Institut f\"ur Astronomie, Universit\"at Wien, T\"urkenschanzstrasse
  17, A-1180, Wien, Austria\\
\and South African Astronomical Observatory, P.O. Box 9, Observatory 
7935,  South Africa\\ 
\and Institut f\"ur Theoretische Physik und Astrophysik, Universit\"at Kiel,
  24098 Kiel, Germany}

\date{Received --; accepted 30-12-2005}

\abstract{We studied the class of DB white dwarf stars, using
re-calibrated UV spectra for thirty four DBs obtained with the IUE satellite.
By comparing the observed energy distributions with model atmospheres, we
simultaneously
determine spectroscopic distances ($d$), effective temperature
($T_{\mathrm{eff}}$), and surface gravities ($\log g$). 
Using parallax measurements and previous determinations of $T_{\mathrm{eff}}$ 
and $\log g$ from optical spectra,
we can study whether the atmospheres of eleven DB stars are consistent 
with pure He or
have a small amount of H contamination. We also report on our
observations of seventeen stars with $T_{\mathrm{eff}}$ close to the DB
instability strip through time series photometry and found them to be non
variable within our detection limits. 
}

\titlerunning{He rich white dwarf stars}
\authorrunning{B. G. Castanheira  et al.}

\maketitle


\section{Introduction}

Among all known white dwarf stars, around 20\% have a helium (He) dominated
atmosphere, and are thus assigned the spectral type DB. 
Most of these stars are believed to be result of the born again or 
a very late He thermal pulse during the early planetary nebula cooling phase 
(e.g. Althaus et al.
2005). In this event,
the residual hydrogen (H) is completely burnt, the star returns quickly to AGB
phase and again to planetary nebula, this time, without H.
As they stars cool down, DBs cross an instability strip, where 
they are seen as
multi-periodic pulsators. 
Beauchamp et al. (1999) determined its boundaries as $27\,800\geq
T_{\mathrm{eff}} \geq 22\,400$\,K from a comparison of their pure He model 
atmosphere grid with ML2/$\alpha$=1.25 to optical spectra, and 
$24\,700\geq T_{\mathrm{eff}} \geq 21\,800$\,K, if undetectable traces of 
hydrogen (H) are allowed in the models. 

The study of the instability strip of the DBs is still a challenge because
of the small number of known pulsators; only seventeen are known to date 
(Nitta et al. 2005). 
Another difficulty is that the determinations of
$T_{\mathrm{eff}}$ and $\log g$ from spectra are degenerate as, 
in general, these two
parameters are correlated.  Working with optical spectra is even more
problematic, as possible contamination with even trace amounts of hydrogen
that are undetected in the spectra can decrease the resulting effective
temperatures by up to 3\,000\,K and $\log g$ by up to 0.05\,dex (Beauchamp
et al. 1999). The uncertainty in $T_{\mathrm{eff}}$ derived from the 
published optical
spectra is thus comparable to the width of the instability strip. 

The DBs have been studied since the 1960s, but
especially after the discovery of a pulsator, GD~358, based on
theoretical predictions (Winget et al. 1982). This star is the brightest
and one of the best studied variable He atmosphere white dwarf (DBV)
stars.  
Because pulsation theory gives detailed predictions of DBV
properties, these stars can be used to study neutrino rates probing the
electro-weak theory (Winget et al. 2004, Althaus \& C\'orsico 2004), the
C($\alpha$,$\gamma$)O cross section (Metcalfe 2003, 2005), and the He$^3$/He$^4$
separation (Wolff et al. 2002, Montgomery \& Winget 2000) which cannot be
achieved in any terrestrial laboratory.
Pulsations in DBs are predicted to exist in a narrow
temperature range, $\sim 3\,000$\,K wide, but it has been difficult to
measure $T_{\mathrm{eff}}$ with sufficient accuracy to determine the
edges of the instability strip.

Considerable interest is focused on the accurate determination of
atmospheric parameters for DB white dwarfs, for yet another reason, the so
called ``DB gap'', where there are no observed DB stars. It occurs between
45\,000 and 30\,000\,K in the cooling sequence (e.g. Hansen \& Liebert
2003). The physical reason of the DB gap is still not
understood. However, many theories attempt to explain why there would
be no DBs within this range of temperature. One possibility is that DBs 
would turn into DAs (white dwarf with pure H atmospheres) by dragging H to 
the surface of the star, blocking
the atmosphere. In this scenario, we expect to find more H in the hot DBs
than in the cooler ones. We also investigate that possibility in this paper,
but we do not confirm this theory.

\section{Fitting the ultraviolet spectra}

To study the DBs as a class and the characteristics of their instability
strip, we used ultraviolet spectra
because they are less affected by possible trace amounts of H that plague the
optical determination of the effective temperature (Beauchamp et al. 1999).
The data we use to determine the distance ($d$), effective temperature
($T_{\mathrm{eff}}$), and surface gravity ($\log g$) are the re-calibrated
ultraviolet spectra for DB stars, obtained with the International Ultraviolet
Explorer (IUE) satellite and published by Holberg, 
Barstow \& Burleigh (2001). 
The spectra were re-calibrated with the New Spectroscopic Image
Processing System (NEWSIPS) data reduction by NASA, and 
in the low-dispersion spectral mode with a resolution of $\sim 6$\,\AA. 

One of the major motivations to use the archive of IUE low-dispersion spectra,
besides it comprising an homogeneous sample, is to work with 
spectra of which the absolute
calibration is based on a synthetic model atmosphere energy distribution for the
white dwarf star G191-B2B (WD~0501+527). The models we fit are the 
same kind used in the flux calibration.

We used a new grid of Koester's model atmospheres, with input physics and
methods similar to those described
in Finley et al. (1997), consisting of models with $T_{\mathrm{eff}}$
from 12\,000\,K to 28\,000\,K, and a step of 500\,K, and $\log g$
from 7.0 to 9.0, with 0.1 dex step. We used two sets of model atmospheres: 
pure He and He contaminated with a small amount of H
$[\log y\equiv \log(N\mathrm{He}/N\mathrm{H})=-3.0]$. This is the 
the upper limit for the amount of H contamination for a star not show 
discernible H lines in the optical spectra, i.e., to be classified as a DB and
not as a DBA.
All models were calculated with
ML2/$\alpha$=0.6 mixing length theory, considering that 
Bergeron et al. (1995) and Koester \& Vauclair (1997) have shown
this choice of mixing length gives consistent results in the UV and optical, for
the DAs. There is no reason to expect the mixing length description to be 
different for DBs.
These models
were used to simultaneously fit $T_{\mathrm{eff}}$, $\log g$ and $d$
to the available IUE spectra.

We calculated the minima in ${\chi}^2$ between the observed spectra and
the models, allowing the three parameters, $T_{\mathrm{eff}}$, $\log g$,
and $d$, to vary. We used the model radii described in Althaus \& 
Benvenuto (1997), available in 
{\tt http://www.fcaglp.unlp.edu.ar/evolgroup/tracks.\\
html}. 

Our determinations of $T_{\mathrm{eff}}$, $\log g$ and the distance
for all DB stars with IUE spectra available are shown in
Table~\ref{table_uv}. In columns 3-5, we show the values derived using pure He 
models,
and in columns 6-8, the same parameters using He/H models.

\begin{table*}
{\small
\begin{tabular}{||cc|c|c|c||c|c|c||}\hline\hline
Name & WD & He $T_{\mathrm{eff}}$ (K) & He $\log g$ & He $d$ (pc) & 
He/H $T_{\mathrm{eff}}$ (K) &
He/H $\log g$ & He/H $d$ (pc)\\ \hline 
G 266-32 & 0000-170 & 16\,000$\pm$600 & 8.50$\pm$0.60 & 39$\pm$12 & 
14\,000$\pm$270 & 8.00$\pm$0.02 & 46$\pm$3 \\
GD~408 & 0002+729 & 14\,000$\pm$40 & 8.50$\pm$0.04 & 26$\pm$1 & 14\,000$\pm$40 &
7.50$\pm$0.07 & 54$\pm$2 \\
Feige 4 & 0017+136 & 19\,000$\pm$170 & 8.00$\pm$0.17 & 61$\pm$5 & 
18\,000$\pm$60 & 7.50$\pm$0.06 & 112$\pm$7 \\
G~270-124 & 0100-068 & 20\,500$\pm$130 & 8.40$\pm$0.23 & 34$\pm$4 & 
19\,000$\pm$100 & 7.00$\pm$0.15 & 69$\pm$5\\
PG~0112+104 & 0112+104 & 27\,000$\pm$110 & 7.50$\pm$0.03 & 136$\pm$2 & 
27\,000$\pm$130 & 8.50$\pm$0.06 & 71$\pm$2 \\
GD~40 & 0300-013 & 15\,000$\pm$420 & 7.50$\pm$0.21 & 104$\pm$11 &
15\,000$\pm$310 & 8.50$\pm$0.16 & 58$\pm$5 \\ 
BPM 17088 & 0308-565 & 21\,500$\pm$190 & 7.70$\pm$0.08 & 58$\pm$2 & 
21\,000$\pm$280 & 8.00$\pm$0.14 & 49$\pm$3 \\
BPM 17731 & 0418-539 & 20\,000$\pm$140 & 8.00$\pm$0.14 & 83$\pm$6 & 
19\,000$\pm$110 & 7.50$\pm$0.06 & 110$\pm$3 \\
BPM 18164 & 0615-591 & 16\,000$\pm$50 & 8.50$\pm$0.17 & 26$\pm$2 & 
16\,000$\pm$40 & 7.00$\pm$0.04 & 64$\pm$1 \\
Ton 10 & 0840+262 & 21\,000$\pm$90 & 7.50$\pm$0.09 & 96$\pm$4 & 18\,000$\pm$100 
& 7.00$\pm$0.15 & 103$\pm$8 \\ 
L748-70 & 0845-188 & 18\,000$\pm$140 & 7.50$\pm$0.14 & 129$\pm$9 & 
18\,000$\pm$60 & 7.50$\pm$0.06 & 135$\pm$4 \\
PG 0853+163$^*$& 0853+163 & 21\,000$\pm$450 & 7.70$\pm$0.18 & 126$\pm$11 & 
20\,000$\pm$650 & 7.50$\pm$0.65 & 137$\pm$44 \\
PG 0948+013 & 0948+013 & 19\,000$\pm$280 & 8.20$\pm$0.19 & 92$\pm$9 & 
18\,000$\pm$150 & 7.00$\pm$0.15 & 173$\pm$13 \\
GD 303 & 1011+570 & 18\,000$\pm$140 & 7.50$\pm$0.14 & 75$\pm$5 & 18\,000$\pm$60 
& 7.50$\pm$0.03 & 78$\pm$1\\
PG 1115+158 & 1115+158& 23\,000$\pm$500 & 8.50$\pm$0.10 & 137$\pm$7 & 
22\,000$\pm$500 & 7.00$\pm$0.25 & 321$\pm$40 \\ 
PG 1149-133$^*$ & 1149-133 & 20\,500$\pm$440 & 7.60$\pm$0.57 & 161$\pm$46 & 
19\,000$\pm$260 & 7.00$\pm$0.19 & 196$\pm$19 \\
PG 1311+129$^*$ & 1311+129 & 26\,500$\pm$450 & 7.70$\pm$0.05 & 249$\pm$6 & 
27\,000$\pm$280 & 7.50$\pm$0.14 & 298$\pm$14 \\
PG 1326-037 & 1326-037 & 21\,500$\pm$290 & 8.40$\pm$0.35 & 81$\pm$14 & 
20\,000$\pm$100 & 8.00$\pm$0.10 & 100$\pm$5 \\
GD 325 & 1333+487 & 16\,000$\pm$40 & 8.20$\pm$0.01 & 34$\pm$0.2 & 
15\,000$\pm$120 & 7.00$\pm$0.06 & 61$\pm$2 \\
PG 1351+489 & 1351+489 & 22\,500$\pm$190 & 7.60$\pm$0.15 & 194$\pm$14 & 
22\,000$\pm$150 & 7.00$\pm$0.07 & 266$\pm$10 \\
PG 1411+218 & 1411+218 & 15\,000$\pm$70 & 7.80$\pm$0.01 & 49$\pm$0.3 & 
14\,000$\pm$70 & 7.00$\pm$0.03 & 66$\pm$1 \\
G 200-39 & 1425+540 & 15\,000$\pm$110 & 7.70$\pm$0.09 & 74$\pm$3 & 
15\,000$\pm$310
& 8.50$\pm$0.31 & 47$\pm$7\\ 
PG 1445+152 & 1445+152 & 21\,500$\pm$120 & 8.40$\pm$0.12 & 91$\pm$5 & 
21\,000$\pm$120 & 8.50$\pm$0.06 & 86$\pm$3 \\ 
PG 1456+103$^*$ & 1456+103 & 24\,000$\pm$190 & 8.50$\pm$0.27 & 110$\pm$15 & 
24\,000$\pm$290 & 9.00$\pm$0.14 & 76$\pm$5 \\
G 256-18 & 1459+821 & 16\,000$\pm$50 & 8.00$\pm$0.03 & 53$\pm$1 & 
15\,000$\pm$310 &
7.00$\pm$0.05 & 83$\pm$2 \\
GD 190 & 1542+182 & 22\,500$\pm$90 & 8.50$\pm$0.11 & 48$\pm$3 & 21\,000$\pm$60 &
7.00$\pm$0.06 & 106$\pm$3 \\
GD 358 & 1645+325 & 24\,500$\pm$130 & 8.50$\pm$0.10 & 29$\pm$1 & 24\,000$\pm$50
& 8.50$\pm$0.03 & 30$\pm$0.4 \\ 
PG 1654+160 & 1654+160 & 25\,000$\pm$550 & 7.50$\pm$0.11 & 237$\pm$13 & 
26\,000$\pm$1\,100 & 7.00$\pm$0.55 & 331$\pm$91\\
L 7-44 & 1708-871  & 23\,000$\pm$610 & 8.30$\pm$0.42 & 55$\pm$12 & 
21\,000$\pm$680
& 7.00$\pm$0.34 & 106$\pm$18 \\
GD 378 & 1822+410 & 17\,000$\pm$60 & 8.20$\pm$0.04 & 39$\pm$1 & 16\,000$\pm$80 &
7.00$\pm$0.08 & 70$\pm$3 \\
L 1573-31 & 1940+374 & 17\,000$\pm$50 & 7.60$\pm$0.05 & 62$\pm$1 & 
17\,000$\pm$40 &
7.00$\pm$0.07 & 86$\pm$3 \\
BPM 26944 & 2034-532 & 17\,000$\pm$390 & 8.50$\pm$0.39 & 34$\pm$7 & 
17\,000$\pm$80 &
7.00$\pm$0.12 & 86$\pm$5 \\
G 26-10 & 2129+000 & 13\,000$\pm$60 & 7.50$\pm$0.06 & 50$\pm$2 & 13\,000$\pm$40 
& 7.50$\pm$0.02 & 54$\pm$1 \\ 
LTT 9031 & 2224-344 & 19\,000$\pm$160 & 7.50$\pm$0.16 & 72$\pm$6 & 
18\,000$\pm$130 &
7.00$\pm$0.13 & 88$\pm$6 \\
\hline \hline
\end{tabular}}
\caption{Atmospheric parameters and distance determined from IUE spectra, using 
pure He models (He) and He contaminated with a small amount of H (He/H) models. 
An asterisk indicates a DBA star, for which our determinations are not
adequate.} 
\label{table_uv}
\end{table*}

\section{Comparison of results}

Our determinations are still degenerate with respect to the contamination of H 
in the He atmosphere. To minimize this effect, we used external measurements,
like: optical spectra, parallax measurements, and V magnitudes,
if available.

\subsection{Distance moduli}

To test the reliability of our spectroscopic distances, we used our
determinations of $T_{\mathrm{eff}}$ and $\log g$, Bergeron's et al.
(2001) absolute magnitude, and the published V magnitude to estimate
the distance moduli. In 
Table~\ref{tabdist} we show the derived distances from this method and the 
distances after cross correlating both solutions.  

In almost all cases, both spectroscopic and magnitude derived distances agree,
even though we used independent model grids.

\begin{table*}
\begin{tabular}{|c||c|c||c|c|c|}\hline\hline
Name & He $d$ (pc) & He/H $d$ (pc) & He $d_{\mathrm{cross}}$ (pc) & 
He/H $d_{\mathrm{cross}}$ (pc) \\ \hline
G 266-32 & $36^{+22}_{-16}$ & 45$\pm$2 & $37^{+13}_{-10}$ & 46$\pm$2 \\
GD~408 & 26$\pm$1 & 51$\pm$2 & 26$\pm$1 & 53$\pm$1 \\
Feige 4  & 81$\pm$7 & 108$^{+4}_{-3}$ & 71$\pm$4 & 110$\pm$4 \\ 
G~270-124  & 35$^{+7}_{-6}$ & 88$^{+10}_{-9}$ & 34$\pm$4 & 78$^{+6}_{-5}$ \\
PG~0112+104 & 144$^{+28}_{-20}$ & 71$^{+3}_{-4}$ & 
140$^{+14}_{-10}$ & 71$\pm$2 \\
GD~40 & 100$^{+20}_{-15}$ & 50$^{+3}_{-8}$ & 102$^{+11}_{-9}$ & 
54$^{+3}_{-5}$\\
BPM 17088 & 61$^{+4}_{-3}$ & 50$\pm$5 & 60$\pm$2 & 50$\pm$3 \\
BPM 17731 & 82$\pm$6 & 109$^{+4}_{-3}$ & 83$\pm$4 & 109$\pm$2 \\
BPM 18164 & 27$\pm$4 & 79$^{+10}_{-9}$ & 27$\pm$2 & 71$^{+5}_{-4}$ \\
Ton 10 & 81$^{+4}_{-2}$ & 104$^{+9}_{-8}$ & 89$^{+3}_{-2}$ & 104$\pm$6 \\
L748-70 & 122$^{+14}_{-12}$ & 122$^{+6}_{-5}$ & 125$^{+8}_{-7}$ & 
128$^{+4}_{-3}$ \\
PG 0853+163$^*$ & 135$^{+18}_{-16}$ & 150$^{+95}_{-56}$ & 131$^{+11}_{-10}$ & 
144$^{+52}_{-36}$ \\
PG 0948+013 & 92$^{+15}_{-13}$ & 205$^{+25}_{-23}$ & 92$^{+9}_{-8}$ & 
189$^{+14}_{-13}$ \\
GD 303 & 119$^{+14}_{-11}$ & 119$^{+7}_{-3}$ & 97$^{+7}_{-6}$ & 
98$^{+3}_{-1}$ \\
PG 1115+158 & 95$^{+8}_{-9}$ & 260$^{+53}_{-45}$ & 116$^{+5}_{-6}$ & 
291$^{+35}_{-30}$ \\
PG 1149-133$^*$ & 178$^{+92}_{-57}$ & 258$^{+42}_{-36}$ & 169$^{+51}_{-37}$ & 
227$^{+23}_{-21}$ \\
PG 1311+129$^*$ & 187$\pm$9 & 218$^{+29}_{-22}$ & 218$\pm$5 & 
258$^{+16}_{-13}$ \\
PG 1326-037 & 82$^{+23}_{-22}$ & 103$^{+7}_{-8}$ & 82$\pm$13 & 101$^{+4}_{-5}$\\
GD 325 & 33$^{+0.2}_{-0.3}$ & 70$\pm$3 & 33$^{+0.1}_{-0.2}$ & 66$\pm$2 \\
PG 1351+489 & 194$^{+21}_{-36}$ & 293$^{+17}_{-15}$ & 194$^{+13}_{-19}$ & 
280$^{+10}_{-9}$ \\
PG 1411+218 & 48$^{+0.5}_{-0.6}$ & 76$\pm$2 & 48$\pm$0.3 & 71$\pm$1 \\
G 200-39 & 69$^{+5}_{-4}$ & 39$^{+11}_{-10}$ & 72$\pm$3 & 43$^{+7}_{-6}$ \\
PG 1445+152 & 77$\pm$7 & 71$\pm$4 & 84$\pm$4 & 78$\pm$2 \\
PG 1456+103$^*$ & 87$\pm$18 & 57$^{+8}_{-7}$ & 98$\pm$12 & 66$\pm$4 \\
G 256-18 & 60$^{+1}_{-2}$ & 111$\pm$7 & 56$\pm$1 & 97$\pm$4\\
GD 190 & 49$^{+4}_{-5}$ & 134$\pm$6 & 49$\pm$3 & 120$\pm$3 \\
GD 358 & 30$\pm$2 & 30$\pm$1 & 30$\pm$1 & 30$^{+0.3}_{-0.4}$ \\
PG 1654+160 & 196$^{+22}_{-16}$ & 302$^{+192}_{-113}$ & 216$^{+13}_{-10}$ & 
317$^{+106}_{-73}$ \\
L 7-44 & 57$^{+14}_{-12}$ & 133$^{+28}_{-22}$ & 56$^{+9}_{-8}$ & 
119$^{+17}_{-14}$ \\
GD 378 & 42$\pm$1 & 90$\pm$6 & 40$\pm$1 & 80$\pm$3 \\
L 1573-31 & 67$\pm$2 & 103$\pm$4 & 64$\pm$1 & 95$\pm$3 \\
BPM 26944 & 34$^{+13}_{-11}$ & 101$^{+10}_{-9}$ & 34$^{+7}_{-6}$ & 
93$^{+6}_{-5}$ \\
G 26-10 & 58$^{+3}_{-2}$ & 58$\pm$1 & 54$\pm$2 & 56$\pm$1\\
LTT 9031  & 78$^{+10}_{-8}$ & 108$^{+12}_{-11}$ & 75$^{+6}_{-5}$ & 
98$\pm$6 \\
\hline \hline
\end{tabular}
\caption{Distance determined from distance modulus (column 2-3) using IUE 
$T_{\mathrm{eff}}$ and $\log g$, 
compared to absolute magnitude models and available V magnitudes. We used both
pure He models (He) and He contaminated with a small amount of H (He/H) models.
The last 2 columns are the distances after cross correlating these values and 
the spectroscopic distances (see values in Table~\ref{table_uv}).
}
\label{tabdist}
\end{table*}

\subsection{Parallax measurements}

For six stars of our sample, parallax measurements are available
(van Altena et al. 2003). 
Comparing these distances with the ones derived spectroscopically, the better
agreement, in general, is the solution derived using pure He models. There are 
two stars, GD 358 and GD 408, for which we could not distinguish the 
atmospheric composition. Feige 4 is an exception,
for which both spectroscopic solutions do not
agree with the published parallax.
However, this is the faintest star in our sample with parallax measurement,
with magnitude close to the limit of the catalog.
In Table~\ref{tab_p}, we show the parallax distance and the best
stellar composition cross correlating the solutions.

\begin{table}
\begin{tabular}{||c|c|c|c||}\hline\hline
Star & $d$ (pc) & Atmosphere\\
\hline
GD 408 & 35$\pm$6 & undetermined \\
Feige 4 & 33$\pm$10 & undetermined \\
GD 325 & 35$\pm$4 & He/H \\
G 200-39 & 58$\pm$13 & pure He \\
GD 358 & 37$\pm$4 & undetermined\\
L 1573-31 & 49$\pm$7 & pure He \\ 
\hline\hline
\end{tabular}
\caption{Using the distance determined by parallax measurements (second column),
we study the best
agreement with our fits, deriving the atmosphere composition (third column).}
\label{tab_p}
\end{table}

\begin{figure}
\centering
\includegraphics[angle=0,width=\linewidth]{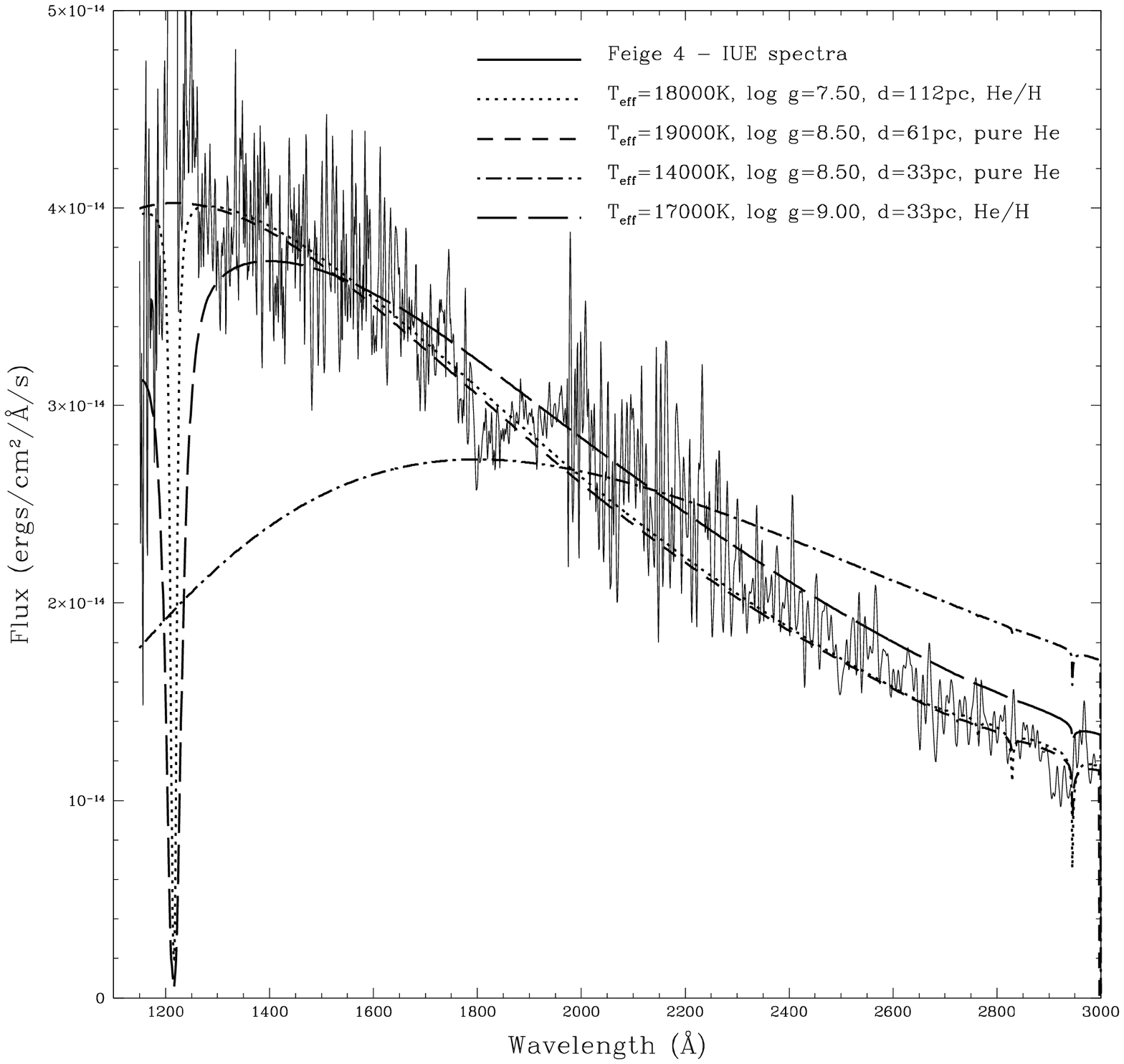}
\caption{IUE spectra obtained for Feige 4 (full line) compared with the best 
models derived leaving all parameters free for a pure He grid (dashed line)
at $d=61$\,pc, with $T_{\mathrm{eff}}=19\,000$\,K and $\log g=8.50$, for a
a DB contaminated with H (dotted line), at $d=112$\,pc, with 
$T_{\mathrm{eff}}=18\,000$\,K and $\log g=7.50$. Using the parallax distance,
($d=33$\,pc) for pure He grid (dotted-dashed line), the best solution is for
$T_{\mathrm{eff}}=14\,000$\,K and $\log g=8.50$, and for a DB contaminated with
H (long dashed line), the atmospheric parameters are 
$T_{\mathrm{eff}}=17\,000$\,K and $\log g=9.00$.}
\label{teste}
\end{figure}

The IUE spectra of Feige 4 (full line)
is shown in Fig.~\ref{teste} in comparison to the
models. The best models derived from the spectra are with 
$T_{\mathrm{eff}}=19\,000$\,K,
$\log g=8.50$, $d=61$\,pc, and pure He (dashed line) and with 
$T_{\mathrm{eff}}=18\,000$\,K, $\log g=7.50$, $d=112$\,pc, and He/H grid
(dotted line). Using the distance derived from parallax, $d=33$\,pc, the best
models are not only much cooler, 
$T_{\mathrm{eff}}=14\,000$\,K for pure He models (dotted-dashed line) and 
$T_{\mathrm{eff}}=17\,000$\,K for He/H models (long dashed line),
but they also do not fit the slope of the observed spectra.
Another argument to claim the parallax measurement is not correct is that this 
star has apparent magnitude V=15.3, too faint for such a large parallax,
unless the radius is extremely small, i.e., high mass, incompatible with the
observed spectra.

\subsection{Comparison with optical spectra results}

Beauchamp et al. (1999) studied the optical spectra of eight known DBVs
together with fifteen other DB and DBA stars with temperatures above
20\,000\,K.  For DBs, including DBVs, they used a pure He atmosphere
composition, or a homogeneous H/He ratio with only traces of
H, at the detection threshold -- defined as that which would produce  barely
visible H$\beta$ or H$\gamma$ features, two lines included in their spectra.
The influence of small, spectroscopically invisible amounts of
H in the DB's atmospheres is an important issue in the definition of the
temperature scale in the optical, because $T_{\mathrm{eff}}$ determined
using He models with small admixture of H are often lower by a few 
thousand of K, than those
determined with pure He models.

\begin{figure}
\centering
\includegraphics[angle=0,width=\linewidth]{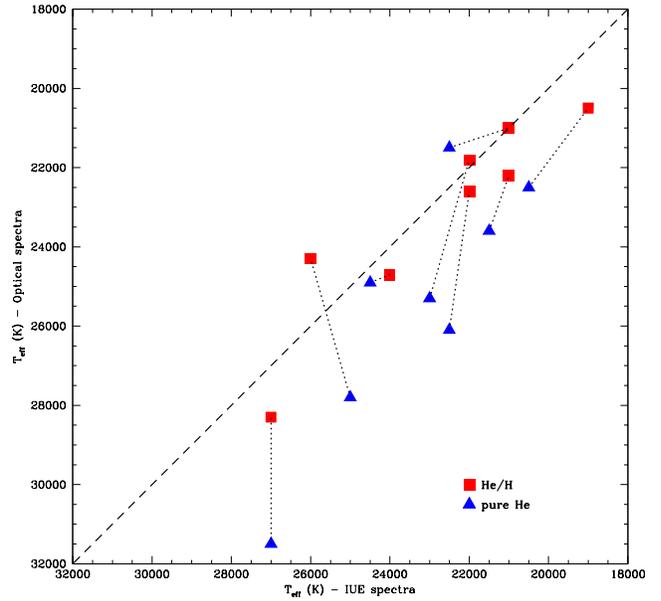}
\caption{Comparison between the UV (x-axis) and optical (y-axis) determinations 
for $T_{\mathrm{eff}}$ using pure He models (blue triangles) and DB models 
contaminated with H (red squares). The dotted lines correspond to the same star.
The dashed line delineates 1:1 correspondence between UV and optical spectra.}  
\label{dif}
\end{figure}

The instability strip Beauchamp et al. (1999) derived from the
analysis of optical spectra contains non-variable stars.
Its $T_{\mathrm{eff}}$ is also uncertain due to the possible presence of trace
amounts of H in the stellar atmospheres.
In Table~\ref{tab_final}, we compare our determination for
$T_{\mathrm{eff}}$ from UV spectra, described in Sect.
2, with
those derived from optical spectra. The optical spectra also give two
solutions, with or without trace H. For seven DB stars, the best agreement in
$T_{\mathrm{eff}}$ in both UV and optical range is for atmospheres 
consistent with a small amount of H instead of none. The exception is the
star GD 358, which has a higher probability of having a pure He atmosphere
in agreement with Provencal et al. (2000) determination of
$\log(N\mathrm{He}/N\mathrm{H}) \leq -5$ for this star.
In Fig.~\ref{dif}, we show a comparison between UV (x-axis) and optical (y-axis)
spectroscopic determinations of $T_{\mathrm{eff}}$, for a pure He atmosphere 
and a He/H atmosphere. The closer a given data point is to the
dashed line (1:1 correspondence between UV and optical spectra), 
the better the solution for the atmosphere composition becomes.
The dotted lines link the two atmosphere determinations
for a given star, showing that He/H atmospheres are more likely for this sample.

GD 358 is the only star in our sample which both parallax measurement and 
optical spectra
determination is available. We cannot distinguish the best atmosphere
composition from the parallax, but a pure He atmosphere is still consistent with
the optical spectra determination. 

For the star GD 190, even though we get a better agreement with the optical 
spectra for a 
contaminated atmosphere, Provencal et al. (2000) obtained an upper limit of 
$\log(N\mathrm{He}/N\mathrm{H}) \leq -6.5$, consistent with a pure He 
atmosphere.

\begin{table*}
\begin{tabular}{|c|c|c||c|c||c||}\hline\hline
Name & UV He & UV He/H & Optical He & Optical He/H  & Atmosphere \\ 
\hline
G~270-124 & 20\,500$\pm$130 & 19\,000$\pm$100 & 22500 & 20500 & He/H\\ 
PG~0112+104 & 27\,000$\pm$110 & 27\,000$\pm$130 & 31500 & 28300 & He/H\\ 
PG 1115+158 & 23\,000$\pm$500 & 22\,000$\pm$500 & 25300 & 21800 & He/H\\ 
PG 1351+489 & 22\,500$\pm$190 & 22\,000$\pm$150 & 26100 & 22600 & He/H\\ 
PG 1445+152 & 21\,500$\pm$120 & 21\,000$\pm$120 & 23600 & 22200 & He/H\\ 
GD 190 & 22\,500$\pm$90 & 21\,000$\pm$60 & 21500 & 21000 & He/H\\ 
GD 358 & 24\,500$\pm$130 & 24\,000$\pm$50 & 24900 & 24700  & He\\ 
PG 1654+160 & 25\,000$\pm$550 & 26\,000$\pm$1\,100  & 27800 & 24300 & He/H \\ 
\hline \hline
\end{tabular}
\caption{Atmospheric parameter determinations from UV spectra
in comparison to those derived by Beauchamp et al. (1999) using optical 
spectra. The last column shows the best agreement in atmosphere composition 
using
both independent determinations.}
\label{tab_final}
\end{table*}

An important consideration is that we fitted all stars using DB models,
never with DBA models. From the IUE spectra, we cannot determine if a star is 
a DBA or 
not. The optical spectra of PG 0853+163, PG 1149-133, PG 1311+129, and 
PG 1456+103 do show H, which has been taken into account by
Beauchamp et al. (1999) by using models with a considerable amount of H.
Our temperatures for DBA stars are
therefore not reliable, but differences in $T_{\mathrm{eff}}$ from our models
with or without trace H are the same order as our uncertainties.

We did not compare $\log g$ values, as their uncertainties are too large
from both UV and optical spectra.

\section{Looking for new pulsators}

Robinson \& Winget (1983) reported a search for pulsating DB
white dwarf stars, classifying twenty nine stars as non-variable.
Expanding this search,
we acquired time-series photometric observations of another thirteen DB
white dwarf stars, which have $T_{\mathrm{eff}}$ close to the edges of the
DB observed instability strip, plus one DA (H atmosphere white dwarf) 
during two observing runs at the South
African Astronomical Observatory (SAAO),  
and four DBs at the Observat\'orio Pico dos Dias (OPD) in other three 
runs, to search for variability. At SAAO, one of us (GH) used the 0.75-m 
telescope in April/May 2000 and
the 1.0-m telescope in December 2001. At both telescopes, a high-speed CCD
photometer (O'Donoghue 1995) was employed. It was operated in full-frame
mode on the 0.75-m telescope with 20-second integrations and 3 - 4
seconds readout during the measurements in 2000, but in frame-transfer
mode with 10-second integrations in 2001. At OPD, we used the 1.6-m
telescope in 1986, with a single channel photometer and 5-second integration
time. We also observed at OPD in 2004, using the 0.6-m telescope and CCD 101,
with 30-second integration, and 7 - 8 seconds readout. No filters were used in 
order to maximize the received light and considering that g-mode pulsations
should have the same phase at different wavelengths (e.g. Kepler et al. 2000). 
We show the observing 
log in Table~\ref{tab2a}.

\begin{table}
\begin{tabular}{||c|c|c|c||}\hline\hline
Star & Run start (UT) & $\Delta$T (hr) & \# points\\
\hline
BPM 17088 & 09/09/86, 05:18 & 1.24 & 890\\
BPM 17731 & 11/09/86, 04:11 & 3.17 & 2282\\
GD 270-124 & 31/10/86, 22:54 & 3.37 & 2423\\ 
WD 0853+163 & 26/04/00, 17:46 & 1.13 & 173 \\
WD 1311+129 & 27/04/00, 21:23 & 1.51 & 230 \\
PG 1445+152 & 28/04/00, 21:18 & 1.33 & 195 \\
PG 0949+094 & 29/04/00, 17:00 & 1.25 & 193\\
PG 1026-056 & 29/04/00, 18:19 & 1.15 & 175\\
L 151-81A & 29/04/00, 23:53 & 1.61 & 250\\
WD 1134+073 & 30/04/00, 18:15 & 1.41 & 213\\
WD 1332+162 & 01/05/00, 18:29 & 1.79 & 260\\
WD 1336+123 & 01/05/00, 20:19 & 1.43 & 222\\
WD 1444-096 & 01/05/00, 21:48 & 1.10 & 168\\
WD 1415+234 & 01/05/00, 22:57 & 0.57 & 88\\
PG 2354+159 & 16/12/01, 18:38 & 1.07 & 386\\
PG 2234+064 & 17/12/01, 18:57 & 0.91 & 326\\
L 7-44 & 15/08/04, 00:33 & 2.61 & 233 \\
\hline\hline
\end{tabular}
\caption{Journal of observations. $\Delta$T is the length of the
corresponding observing run.}
\label{tab2a}
\end{table}

\begin{figure}
\centering
\includegraphics[angle=0,width=10cm]{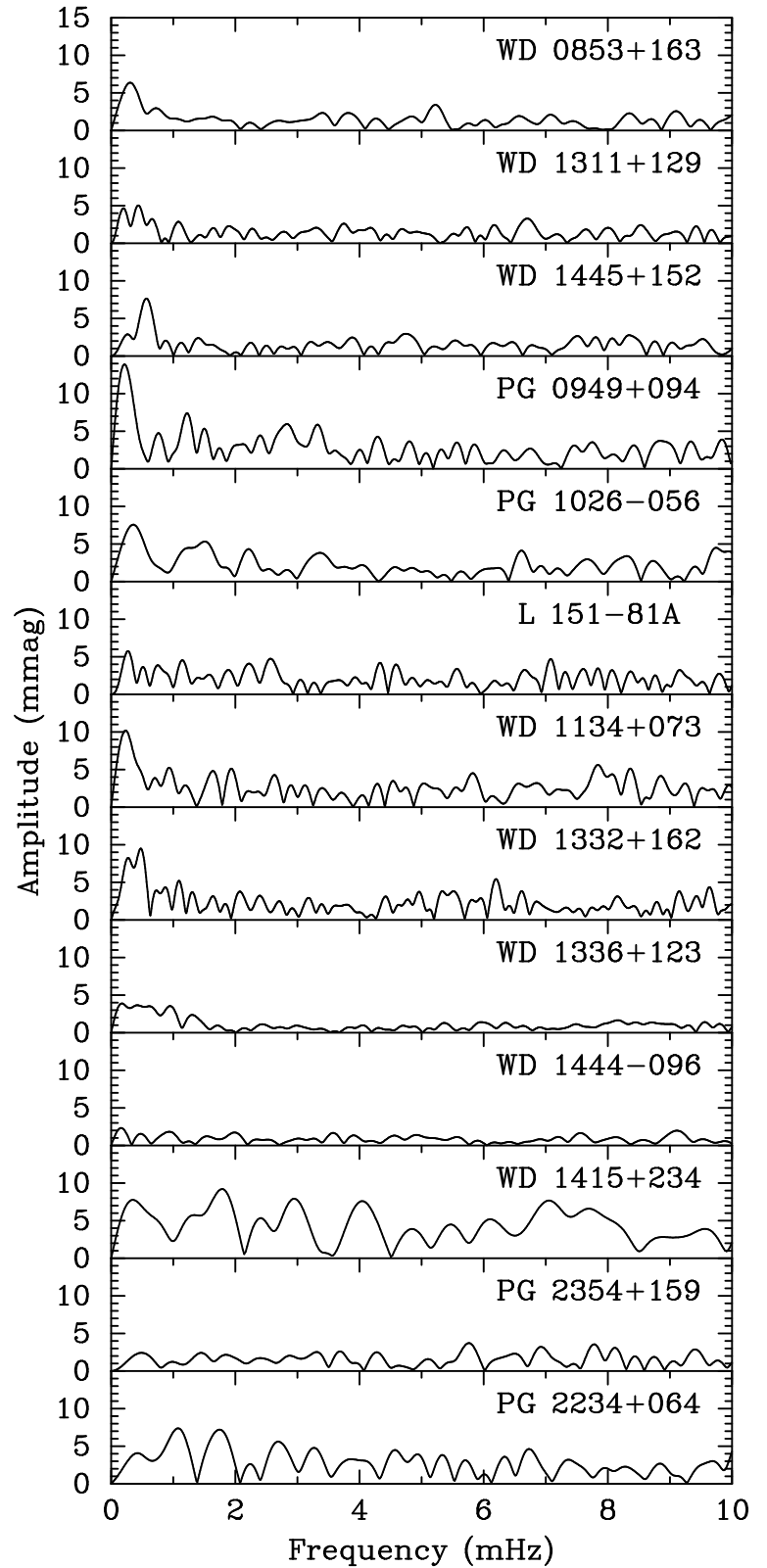}
\caption{Fourier amplitude spectra of the null results of a search for pulsation
among DB white dwarf stars.}
\label{limits}
\end{figure}

We reduced the SAAO CCD data with the standard software for this instrument, and
carried out photometry by using the program MOMF (Kjeldsen \& Frandsen
1992) which uses a combined approach of PSF fitting photometry and
aperture corrections on the star-subtracted frames, giving optimal
results. Fourier amplitude spectra of the resulting light curves are shown in
Fig.~\ref{limits}. For the OPD runs, the detection limits are 3\,mma for 
BPM 17088 and GD 270-124, 2\,mma for BPM 17731, and 1.4\,mma for L 7-44.

All the seventeen stars are constant within our detection limit. 
The detection
limits are satisfactory for all stars except PG 0949+094, WD 1415+234 (run
terminated by cloud) and PG 2234+064, which should be re-observed. WD
1445+152 may also require some additional observations; the highest peak
in its amplitude spectrum is somewhat outside the typical range for
pulsating white dwarf stars but we cannot rule out that it is intrinsic to 
the star
from the present data. We also note that we could not detect
variability of the DA white dwarf L 151-81B, but our detection limit ($\sim$ 8
mma) is poor. On the other hand, we suspect that the star 
2MASS 14581310-6317340, 
($\sim$ 8 arc-seconds East of L 151-81AB) is a $\delta$ Scuti star, with a
1.3-hr period and 23 mmag semi-amplitude.

The variability classification of DB stars 
is shown in Table~\ref{var},
where V is used for variables, NV for non-variables, and NO for not observed
for variability reported.

\begin{table}
\begin{tabular}{|c|l|}\hline
& Stars\\ \hline
V & PG 1115+158, PG 1351+489, PG 1456+103, \\ 
& GD 358, PG 1654+160\\
NV & Feige 4, G270--124, PG 0112+104, GD 40, \\
& BPM 17731, Ton 10, PG 0853+163, GD 303, \\
& PG 1311+129, GD 325, PG 1445+152, G 256-18,\\ 
& GD 190, L 7-44, GD 378, G 26-10, LTT 9031 \\
& BPM 17088 \\
NO & G 226-32, GD 408, BPM 18164, L 748-70, \\
& PG 0948+013, PG 1149-133, PG 1326-037, \\
& PG 1411+218, G 200-39, L 1537-31, BPM 26944\\ \hline
\end{tabular}
\caption{Variability classification of the DB stars in our sample. The V
is used for variables, NV for non-variables and NO for not observed for
variability.} 
\label{var}
\end{table}

\begin{figure}
\centering
\includegraphics[angle=-90,width=\linewidth]{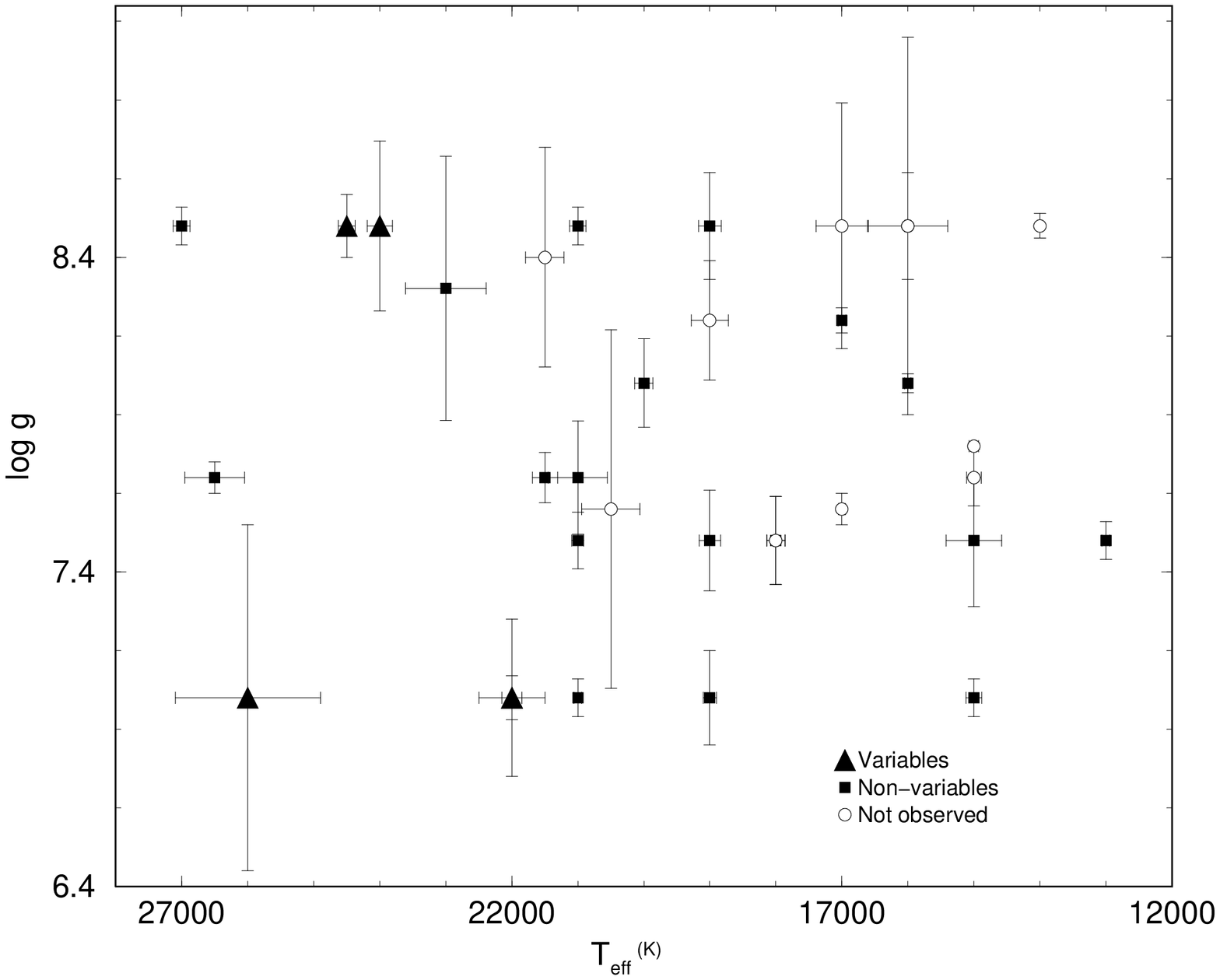}
\caption{DB instability strip using pure He models for the stars for which
we cannot determine
atmosphere composition: variables (filled triangle), non-variables
(filled squares), and not observed for variability (open circles). There are 2 
stars close to
the instability strip that have not been observed for variability.}
\label{inst}
\end{figure}

\begin{figure}
\centering
\includegraphics[angle=-90,width=\linewidth]{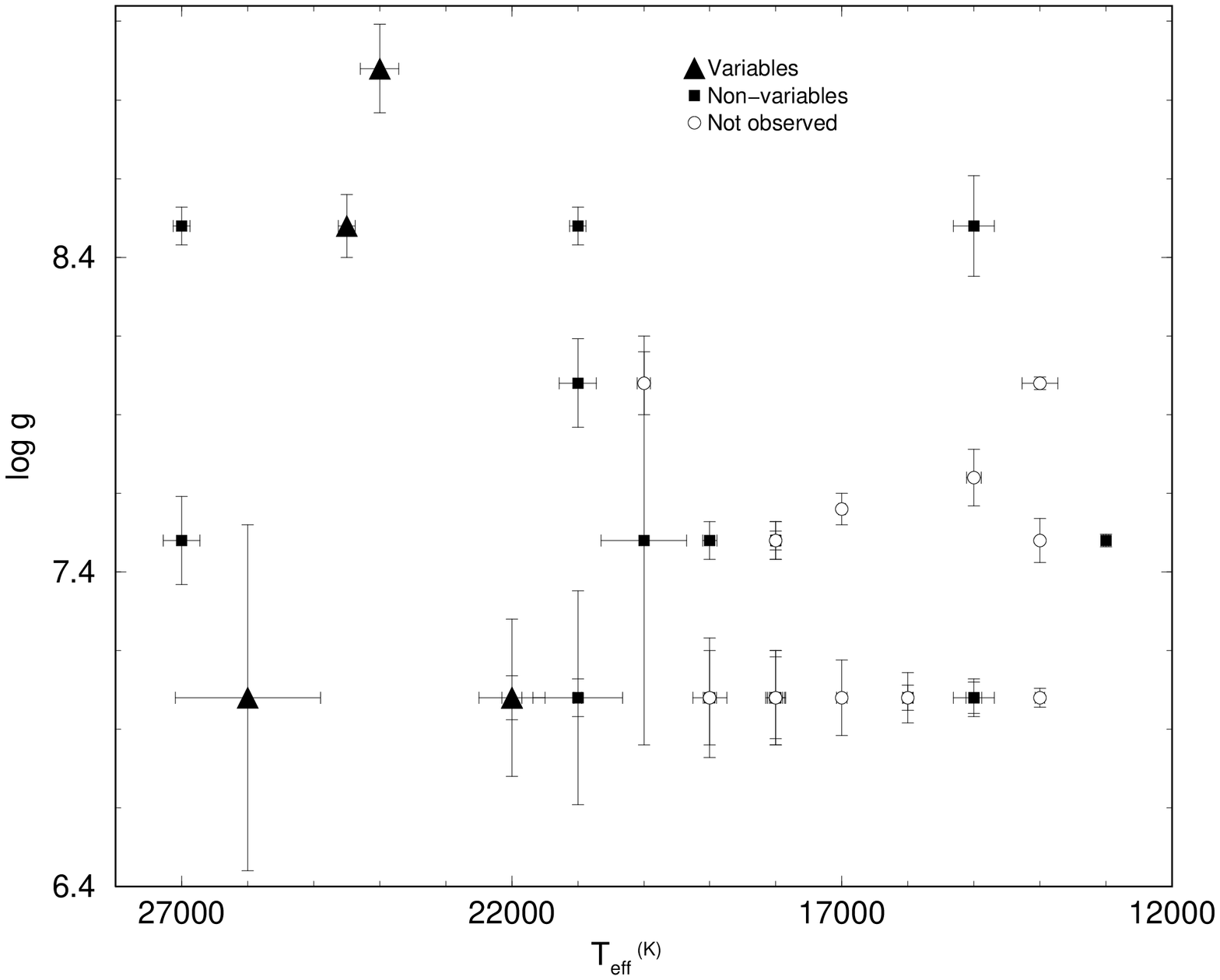}
\caption{DB instability strip using H contaminated He models for the stars 
for which we cannot
determine atmosphere composition: variables (filled triangle), 
non-variables
(filled squares), and not observed for variability (open circles). There are 2 
stars close to
the instability strip that have not been observed for variability.}
\label{inst2}
\end{figure}

Having derived the physical parameters from ultraviolet spectra, and the
atmosphere composition for thirteen stars, we are
ready to determine the DB instability strip for this homogeneous
sample.  In Fig.~\ref{inst} and Fig.~\ref{inst2}, 
we show the final determination
for $T_{\mathrm{eff}}$ and $\log g$ for 
variables (filled triangle),
non-variables (filled squares), and so far not observed by time series 
photometry
(open circles) DB stars. This diagram shows 
that DBs pulsate in a well-defined temperature range, from 26\,000 $\geq
T_{\mathrm{eff}} \geq$ 22\,000\,K. For the stars which we could not determine
their atmosphere composition, we used both pure He models and He/H models, 
respectively. 
There is a 97\% chance that the DB instability strip 
contains only variable stars. Even if the error bars in 
$T_{\mathrm{eff}}$ were three times larger, there is only a 4\% probability
of contamination. This probability was calculated by adding the
probability of all variables that fall inside the instability strip and all
non-variables outside, using Gaussian distributions for our $T_{\mathrm{eff}}$
determinations only (no consideration for $\log g$). 
There are still stars close
to the instability strip that have not been searched to our knowledge 
for variability, and which are crucial for the study of the instability strip.

However, Beauchamp et al.'s (1999) optical spectra fitting found 
non-variables
inside the instability strip. In this sense, for a true determination
of the DB instability strip it is necessary to fit the optical and UV spectra
simultaneously, to analyze the possible differences to to convection 
prescription.

\section{Concluding remarks}                                  

We used model atmospheres with
ML2/$\alpha$=0.6 to derive atmospheric parameters ($T_{\mathrm{eff}}$ and 
$\log g$) and distances
for thirty four DB stars with available IUE re-calibrated spectra.
Our model grid fit well the spectra.
Another important conclusion is that atmospheric 
contamination with H is
not directly proportional to $T_{\mathrm{eff}}$ 
for DB stars, based on our determination for eleven stars, which has been
a suggestion to explain the DB gap by convection dragging H upwards.
We also find no DB stars inside the DB gap.

\begin{acknowledgements}                                      
Financial support: CAPES/UT grant, CNPq fellowship.
\end{acknowledgements}                                        
                                                              
%
%

\end{document}